\documentclass[reprint,preprintnumbers,amsmath,amssymb]{revtex4-1}
\usepackage{graphicx}
\usepackage{dcolumn}
\usepackage{bm}
\usepackage{color}
\usepackage{hyperref}
\usepackage{verbatim}
\usepackage{subfig}

\begin{document}

\begin{abstract}
\noindent We present a theoretical study of the energetics of thin nematic shells with two charge one-half defects and one charge-one defect. We determine the optimal arrangement: the defects are located on a great circle at the vertices of an isosceles triangle with angles of  $66^{\circ}$ at the charge one-half defects and a distinct angle of $48^{\circ}$, consistent with experimental findings. We also analyse thermal fluctuations around this ground state and estimate the energy as a function of thickness. We find that the energy of the three-defect shell is close to the energy of other known configurations having two charge-one and four charge one-half defects. This finding, together with the large energy barriers separating one configuration from the others, explains their observation in experiments as well as theirlong-time stability.
\end{abstract}

\title{Spherical nematics with a threefold valence}

\author{Vinzenz Koning$^{\dag}$, Teresa Lopez-Leon$^{\dag\dag}$, Alexandre Darmon$^{\dag\dag}$
Alberto Fernandez-Nieves$^{\dag\dag\dag}$ and V. Vitelli$^{\dag}$}
\affiliation{$^{\dag}$Instituut-Lorentz
for Theoretical Physics, Leiden University, Leiden NL 2333 CA, The Netherlands. \\
$^{\dag \dag}$EC2M, UMR Gulliver CNRS-ESPCI 7083 – 10 Rue Vauquelin, F-75231 Paris Cedex 05, France \\
$^{\dag\dag \dag}$School of Physics, Georgia Institute of Technology, Atlanta, Georgia 30332, USA. 
}

\email{E-mail: koning@lorentz.leidenuniv.nl}

\date{\today}

\maketitle

\section{Introduction}

\begin{figure}[h]
\centering
\includegraphics[width=\columnwidth]{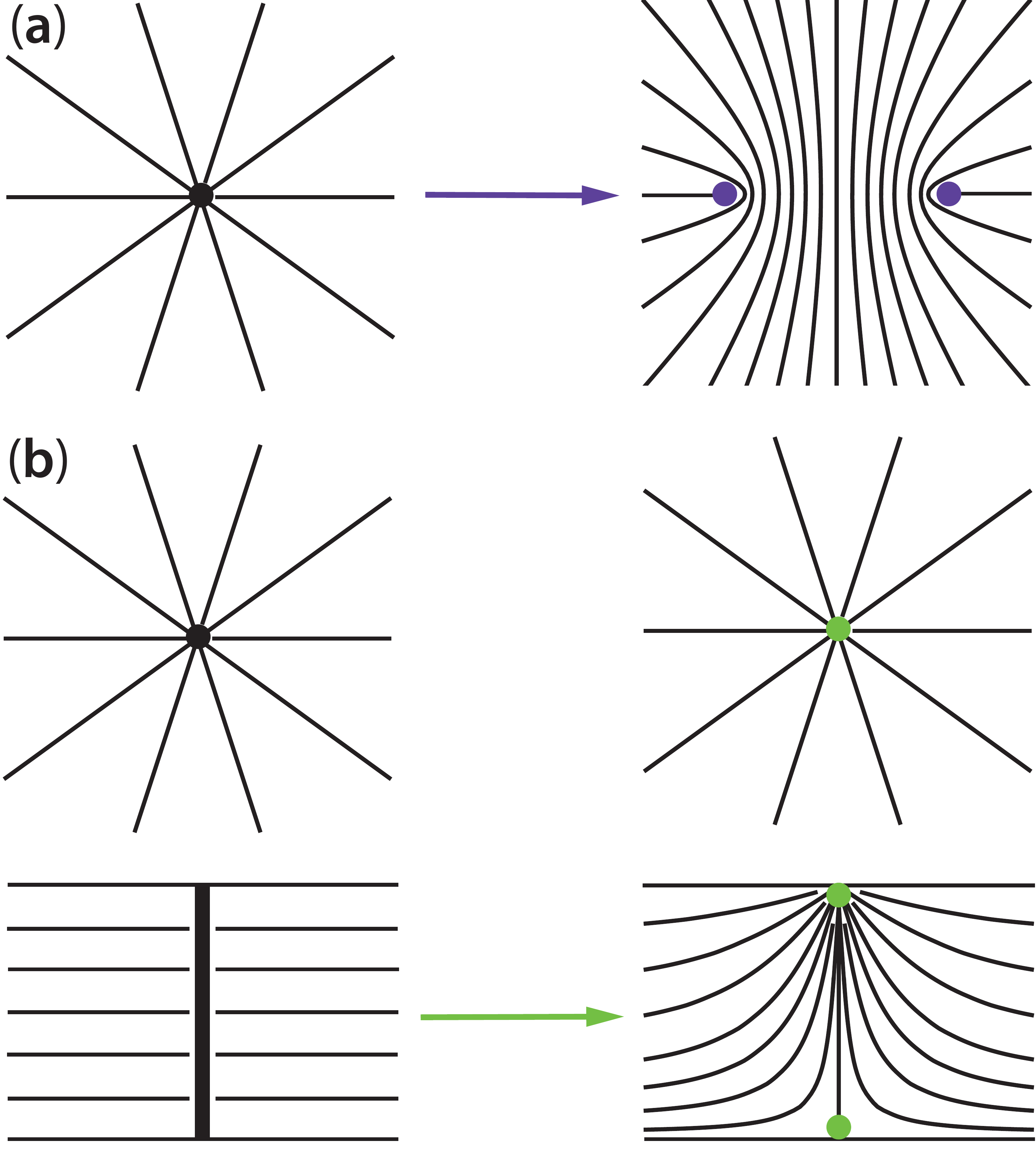} 
\caption{(a) In a two-dimensional nematic a $s=1$ topological defect (black dot in left panel) can lower its elastic energy by splitting into two $s=1/2$ defects (purple dots in right panel). (b) A singular line (left panel) spanning the shell with a winding number of one at the boundaries is topologically and energetically unstable. The singular core is indicated by a black dot in the top view shown in the top panel and by the vertical bold line in a cut shown in the bottom panel. One way of reducing the elastic energy escaping in the third (vertical) dimension (right panel), thereby leaving a point defect (green dot), called boojum, on each boundary. \label{fig:escape_and_separation}}
\end{figure}

One of todays major drives in condensed matter physics is the assembly of mesoscale particles into complex structures \cite{Glotzer:2007fk}.  
By creating anisotropy in the interparticle interactions, one can increase the complexity and functionality of these structures.
A proposed way to achieve anisotropic interactions is by coating a spherical particle or droplet with an orientationally ordered phase \cite{2002NanoL...2.1125N}. The topology of the sphere enforces defects in the coating. Since these defects are very distinct regions on the sphere, they are suitable for the attachment of linkers acting as bonds between the particles. For the case of a vector order parameter, topology requires two defects, creating a particle with two binding sites. Exploiting this, de Vries \textit{et al.} succesfully assembled chains of such divalent nanoparticles \cite{2007Sci...315..358D}.
Nematic rather than vector order allows for defects of charge one-half, referring to the 180 degrees rotation experienced by the local average orientation of the nematic molecules,  $\mathbf{n}$, when encircling the defect. In fact, it is energetically favourable for defects of charge one to split into two charge one-half defects (Fig. \ref{fig:escape_and_separation}a). Nematic order on the sphere has four topological defects of charge one half in its ground state, such that the sum of all charges is equal to $2$, the Euler characteristic of the sphere, as demanded by the Poincare-Hopf theorem. Their mutual repulsion drives them as far away from each other as possible: at the vertices of a regular tetrahedron \cite{1992JPhy2...2..371L}.
Thus, chemical functionalisation of the defects in the ground-state of two-dimensional nematic liquid crystal on the sphere would thus result in the diamond structure \cite{2002NanoL...2.1125N}.  In the decade that followed the conception of this idea, a vast amount of theoretical and numerical work was performed; it included studying the effect of elastic anisotropies, external fields, sphericity or shape, and thickness and thickness inhomogeneity of the nematic film \cite{2006PhRvE..74b1711V,2007PhRvL..99o7801F,PhysRevLett.100.197802,2008SMat....4.2059B,2008JChPh.128j4707B,2008PhRvL.101c7802S,B917180K,2011NatPh...7..391L,C0SM00378F,PhysRevE.85.061701,PhysRevLett.108.207803,PhysRevLett.108.057801,PhysRevE.85.061710,2012PhRvE..86b0705S,PhysRevE.86.020703,Napoli201366,C3SM27671F,C3SM50489A,C2SM07384F,PhysRevE.86.011709,PhysRevE.88.012508,PhysRevE.89.052504, PhysRevE.80.051703, PhysRevLett.99.017801, PhysRevLett.108.017801}. These works were mainly focussing on spherical nematic shells with four defects with a charge of one-half. Experimental investigations on nematic shells generated by trapping a water droplet inside a nematic droplet, however, have revealed the existence of a much wider variety of defect structures besides the regular tetrahedral defect arrangement\cite{2007PhRvL..99o7801F, 2011NatPh...7..391L, 2011PhRvL.106x7802L,2011PhRvL.106x7801L,Lopez-Leon:2011fk, 2012PhRvE..86b0705S,C3SM27671F}, even with a valence number different from four\cite{2007PhRvL..99o7801F, 2011NatPh...7..391L, C3SM27671F}. There exist divalent configurations in which instead of four charge half defect lines spanning the shell, there are two pairs of point defects, called boojums, residing on the boundary surfaces. They arise because the thickness of the nematic coating is nonzero: the elastic energy of a singular line with a winding number of one at the boundary is reduced by escaping in the third dimension, as illustrated in Fig. \ref{fig:escape_and_separation}b. This route thus forms an alternative to the splitting into $s=1/2$ lines spanning the shell. 

Surprisingly, also structures containing both boojums as well as charge one-half disclination lines have been observed in experiments \cite{2007PhRvL..99o7801F, 2011NatPh...7..391L}. These defects structures have threefold valence yet they are still consistent with Poincare-Hopf's theorem, because the total topological charge of the defects at the boundary is $1+1/2+1/2=2$, the Euler characteristic of the sphere. Again, also this energetically stable defect configuration arises because of the finite thickness of the nematic coating. If the shell is strongly inhomogeneous in thickness, experiments show that the defects cluster in the thin part of the shell \cite{2007PhRvL..99o7801F, 2011NatPh...7..391L}. In simulations, the trivalent state has also been observed for inhomogeneous shells \cite{PhysRevE.88.012508}. The defects were found to be positioned at the vertices of an isosceles triangle, with the boojums located in the thickest region of the shell and the $s=1/2$ disclination lines in the thinnest hemisphere.

In this report, we will study theoretically the defect separations, energetics and fidelity of the bonds in homogeneous spherical nematic shells with three-fold valence. We will make a comparison with divalent and tetravalent shells and find the optimal valency as a function of shell thickness as well as the energy barrier between shells of different valency. We will compare these results to experiments in which we decrease the thickness inhomogeneity.

\section{Trivalent ground state}

\begin{figure}[t!]
\centering
\includegraphics[width=\columnwidth]{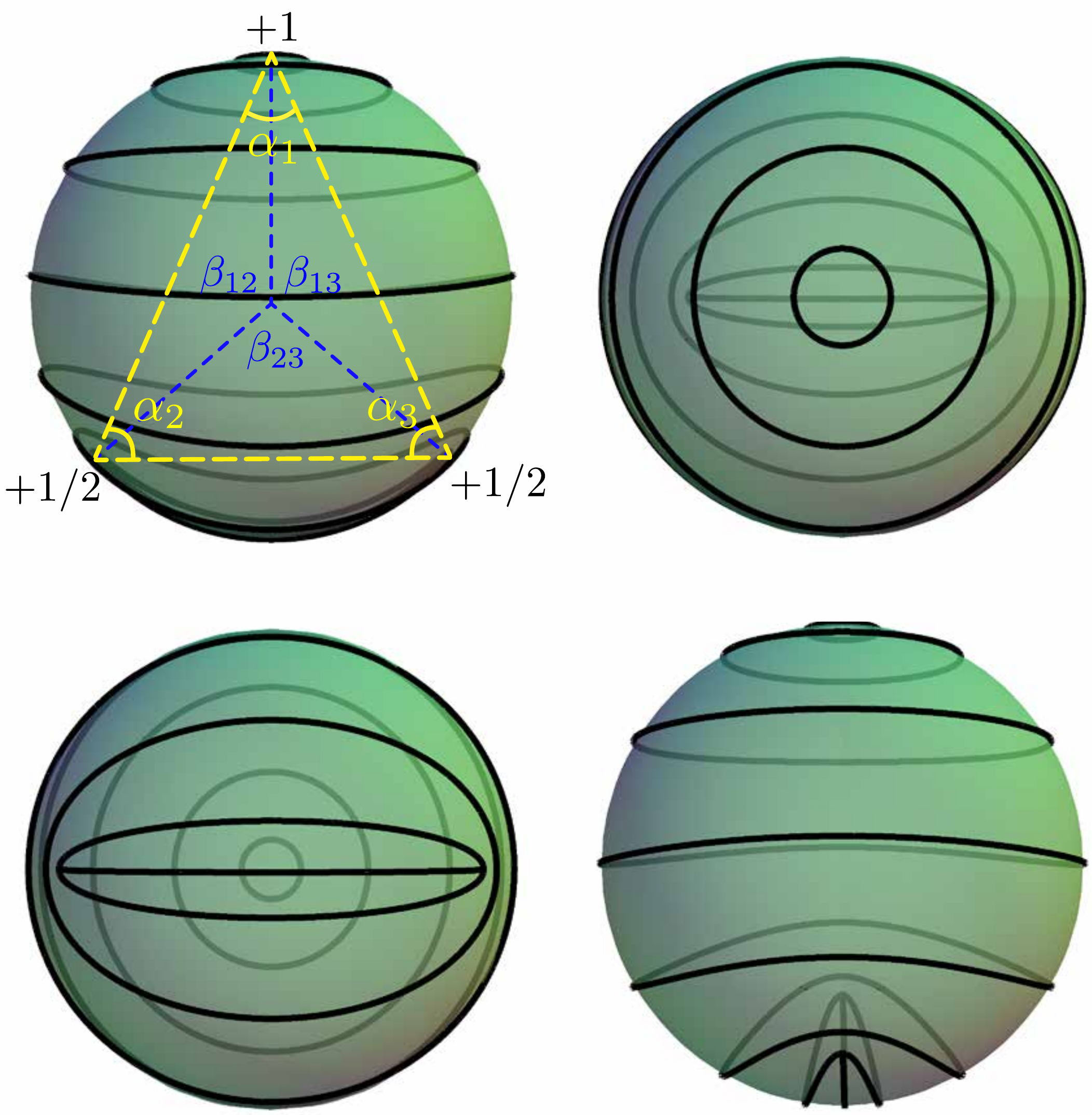} 
\caption{Four views on the bend texture of the director field on the sphere containing a $+1$ defect and two $+1/2$ defects arranged in an isosceles triangle with $ \beta_{12} = \beta_{13} \approx 132^{\circ}$, $\beta_{23} \approx 96.4^{\circ}$, $\alpha_1  \approx 48^{\circ}$ and $\alpha_2 =  \alpha_3 \approx 66^{\circ}$. The defects lie on a great circle.}
\label{fig:splay_ground_state}
\end{figure}

\begin{figure}[t!]
\centering
\includegraphics[width=\columnwidth]{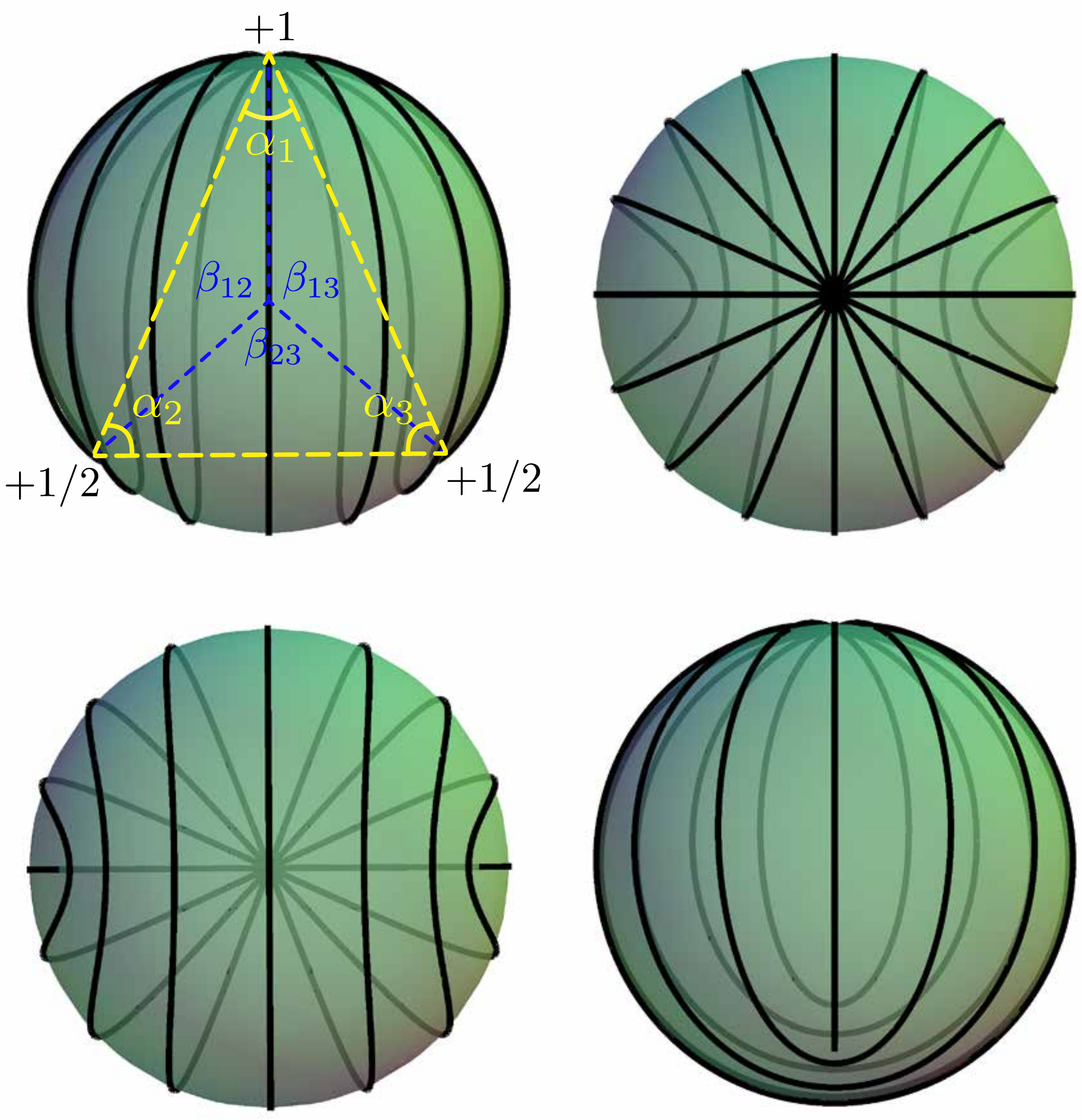} 
\caption{Four views on the splay texture of the director field on the sphere containing a $+1$ defect and two $+1/2$ defects arranged in an isosceles triangle with $ \beta_{12} = \beta_{13} \approx 132^{\circ}$, $\beta_{23} \approx 96.4^{\circ}$, $\alpha_1  \approx 48^{\circ}$ and $\alpha_2 =  \alpha_3 \approx 66^{\circ}$. The defects lie on a great circle.}
\label{fig:bend_ground_state}
\end{figure}

The free energy of a thin curved nematic film is

\begin{equation}
\label{eq:free_energy_curved}
F= \frac{1}{2} \int dA \left[k_1 \left( D_i n^i \right)^2
+ k_3 \left( D_i n_j - D_j n_i \right) \left( D^i n^j - D^j n^i \right)\right]
\end{equation}

\noindent with $k_1$ and $k_3$ the two-dimensional splay and bend elastic constants and $D_i$ is the covariant derivative. Eq. \eqref{eq:free_energy_curved} can be recast in terms of defect separation rather than the director field $\mathbf{n}$. For a spherical nematic, the elastic energy in the one-constant approximation $k=k_1=k_3$ reads

\begin{equation}
\label{eq:free_energy_defect}
F = - \pi k  \sum_{i <  j}^Z s_i s_j \log \left( 1- \cos \beta_{ij} \right) + \sum_i^Z E_i\left(R\right)
\end{equation}

\noindent where $s_i$ is the topological charge of defect $i$, $\beta_{ij}$ is the angular distance between defects $i$ and $j$, and $Z$ is the number of defects or valence number. The self-energy $E_i\left(R\right)$ is given by

\begin{equation}
E_i\left(R\right) = \pi K s_i^2 \log\left( \frac{R}{a}\right) + E_c,
\end{equation} 

\noindent where $R$ is the radius of the sphere and $a$ is a small scale cut-off preventing a divergence of the energy. $E_c$ represents a core energy, which depends on the details of the microscopic interactions. The self-energy is responsible for the splitting of $+1$ defect in an ideal two-dimensional nematic, because of its proportionality with $s_i^2$. The other term in eq. \eqref{eq:free_energy_defect} describes the repulsion between like-charged defects. We wish to find the optimal location for the defects in a thin homogeneous shell given that $s_1=1$, $s_2=\frac{1}{2}$ and $s_3=\frac{1}{2}$. This requires minimising the interaction term of the free energy.  We minimise the interaction energy with respect to three independent variables, namely $\beta_{12}$, $\beta_{13}$ and the angle, $C$, subtended by the two curved triangular sides (circular arcs) meeting at the charge one defect. If we apply the law of cosines on the sphere:

\begin{equation}
\cos \beta_{23} = \cos \beta_{12} \cos \beta_{13} + \sin \beta_{12} \sin \beta_{13} \cos C,
\end{equation}

\noindent we can eliminate $\beta_{23}$ in favour of $C$ in the free energy, and demand $\frac{\partial F}{\partial \beta_{12}}=\frac{\partial F}{\partial \beta_{13}}=\frac{\partial F}{\partial C}=0$. From the latter equation, $\frac{\partial F}{\partial C}=0$, we obtain $C = \pi$, implying that the defects lie on a great circle (see Figs. \ref{fig:splay_ground_state} and \ref{fig:bend_ground_state}). There is always a circle that can be drawn through three points on a sphere; the maximal radius of this circle reflects the repulsive nature of the defects. With some straightforward algebra the other two equations, $\frac{\partial F}{\partial \beta_{12}}$ and $\frac{\partial F}{\partial \beta_{13}}=0$, then lead to 

\begin{align} 
 \beta_{12} &= \beta_{13} = \pi - \arccos \frac{2}{3} \approx 0.73 \pi \approx 131.8^{\circ} \\ 
 \beta_{23} &= 2 \arccos \frac{2}{3} \approx 0.54 \pi \approx 96.4^{\circ}
\end{align}

We thus find that the defects are located at the vertices of an isosceles triangle rather than equilateral triangle, shown in Figs. \ref{fig:splay_ground_state} and \ref{fig:bend_ground_state}. This less symmetric configuration arises because of the asymmetry in the magnitude of the charges of the defect: the two $+1/2$ defects repel each other less strongly than a charge one and charge one-half such that $\beta_{12}$ and $\beta_{13}$ are larger than $\beta_{23}$. This is in marked contrast with the regular tetrahedral configuration in which all the defects are equidistant, because all four charges are indistinguishable. The fact that $s_2$ and $s_3$ are of equal magnitude is still reflected in the equal length of two of the sides ($\beta_{12}=\beta_{23}$) of triangle. Perhaps surprisingly, the distance between two charge one-half defects is smaller in the trivalent state than in the more `crowded' tetravalent state. The surface angles of the flat triangle can be found be simple trigonometry: by realising that the triangle formed by two defects and the centre of the sphere is also an isosceles triangle (of which two sides have a length equal to the radius) we obtain

\begin{align} 
\alpha_1 &= \pi - \beta_{12} = \arccos \frac{2}{3}  \approx 48.2^{\circ} \\
\alpha_2 &=  \alpha_3 = \frac{\beta_{12}}{2} = \frac{\pi}{2}-\frac{\arccos\frac{2}{3}}{2} \approx 65.9^{\circ} 
\end{align}
\newline

Given the defect locations the energy-minimising director field can be found by means of a stereographic projection of the planar solution. The bend texture is displayed in Fig. \ref{fig:bend_ground_state}.  Rotating this director field over an angle $\alpha$ yields the same free energy in the one-constant approximation. The splay texture (Fig. \ref{fig:splay_ground_state}) corresponds to $\alpha=\pi/2$. We note that the escape in the third dimension, in which the singular region is distributed over a larger distance of the order of the thickness, occurs in shells of finite thickness and is somewhat different than the problem of three point defects in a two-dimensional nematic solved above. However, we expect that the defect separations will be marginally affected as long as the thickness is small compared to the radius.

\begin{figure}[t!]
\centering
\includegraphics[width=1.02\columnwidth]{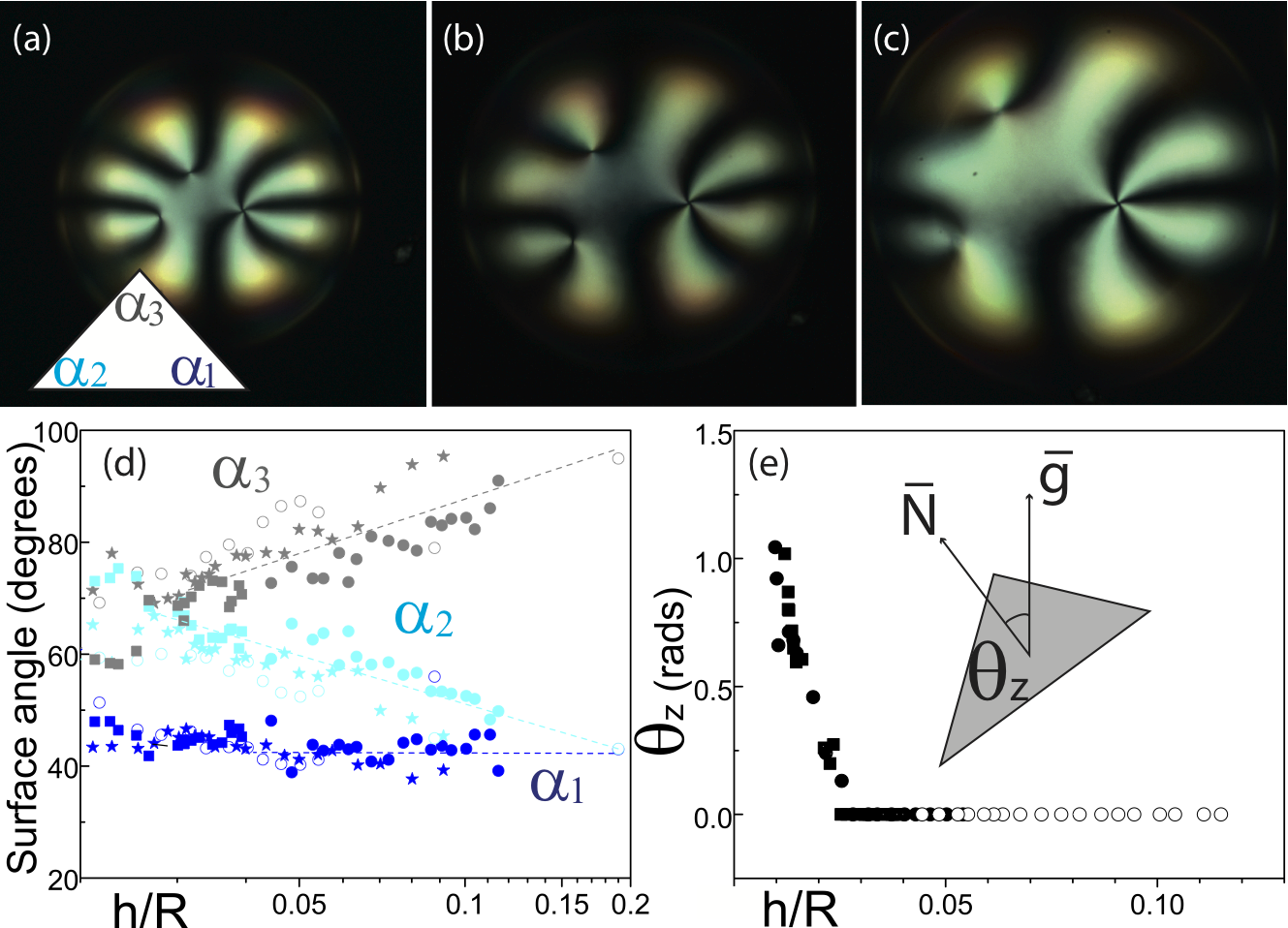} 
\caption{(a-c) Cross-polarized images of a shell at different stages of a swelling process, which makes the shell more homogeneous in thickness. The shell has three topological defects: one defect of charge $+1$, characterized by four black brushes, and two defects of charge $+1/2$, characterized each one by two black brushes. The three defects depict a triangle whose angles, $\alpha_1$, $\alpha_2$ and $\alpha_3$, change as the shell swells, see evolution from left to right. The exact values of $\alpha_1$, $\alpha_2$ and $\alpha_3$ as a function of the shell average thickness $h/R$ are shown in (d). The triangle depicted by the defects is initially oriented perpendicularly to the gravitational direction, but it tilts off as the shell swells, as shown in (d), where $\theta_{z}$ stands for the angle between gravity $\bar{g}$ and the flat triangle normal $\bar{N}$.
\label{fig:experiments1}}
\end{figure}

To test the theoretical expectations, we generate nematic shells using microfluidics \cite{2007PhRvL..99o7801F}. The shells produced with this method are double emulsions where an inner aqueous droplet is contained inside an outer liquid crystal droplet which is, in turn, dispersed in an aqueous solution. We use salt to establish the osmotic pressures of both the inner droplet and the continuous phase and thus the osmotic pressure difference between them. The stability of the emulsion is guaranteed by the presence of polyvinyl alcohol (PVA), which also enforces planar anchoring of the liquid crystal, 4-Cyano-4'-pentylbiphenyl ($5$CB), at both the inner and outer interfaces. The shells are heterogeneous in thickness due to buoyancy effects; they are thinner at the top and thickest at the bottom \cite{2007PhRvL..99o7801F}. Hence, the shell thickness gradually increases from the top to the bottom of the shell. Typical values of the outer radius, $R$, and thickness of the shell, $h$, are in the ranges $[20,60] \mu m$ and $[1,10] \mu m$, respectively. In this shell geometry, the defects are confined to the top, as shown in Fig. \ref{fig:experiments1}(a), since this reduces the Frank free energy \cite{, 2011NatPh...7..391L}.

\begin{figure}[h]
\centering
\includegraphics[width=\columnwidth]{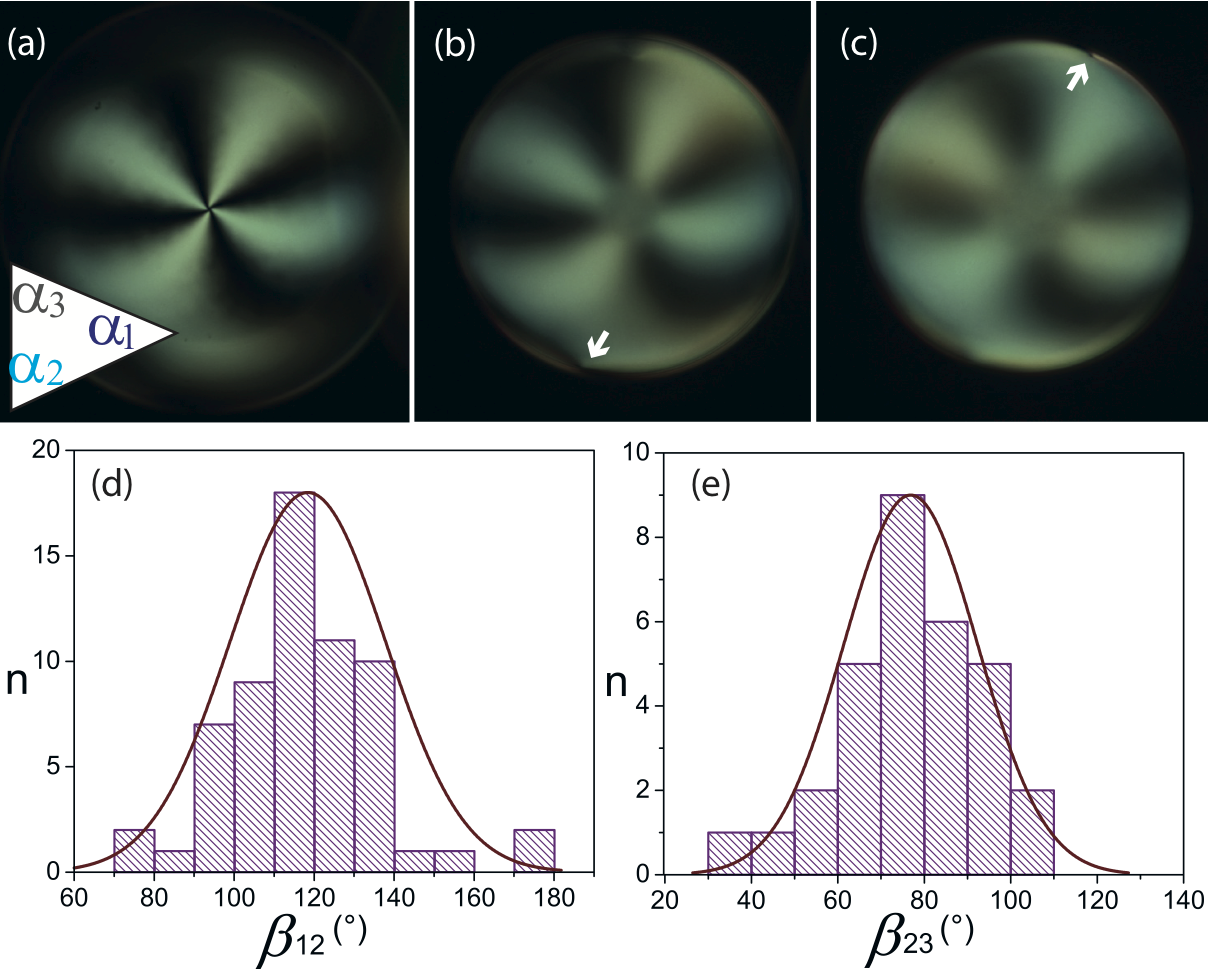} 
\caption{Defect configuration of a very thin and almost homogeneous trivalent nematic shell. (a-c) Cross polarised images of the shell at different focal planes: the $+1$ defect is in an upper plane shown in (a), while the $+1/2$ defects are at lower planes, see arrows in (b) and (c). (d-e) Histograms of the angular distances between defects. The three defects depict an isosceles triangle where the two equal sides correspond to the distance between the +1 defect and each of the +1/2 defects, $\beta_{12}= (118\pm19)^\circ$, and the unequal side corresponds to the distance between the two $+1/2$ defects, $\beta_{23}= (77\pm15) ^\circ$.\label{fig:experiments2}}
\end{figure}

As a result of the imposed osmotic pressure differene between the inner droplet and the continuous phase, the inner droplet swells while the shell becomes thinner and more homogeneous. This happens quasi-statically \cite{, 2011NatPh...7..391L}. We then monitor the evolution of the defects throughout the process. We find that the defects progressively spread, as shown in Figs. \ref{fig:experiments1}(a-c). Interestingly, the shape of the triangle defined by the positions of the three defects changes shape as this happens; the surface angles $\alpha_1$, $\alpha_2$ and $\alpha_3$ progressively evolve as the shell becomes thinner and more homogeneous, as shown in Fig. \ref{fig:experiments1}(d). For shells with $h/r\lesssim 0.03$, the surface angles reach values that are consistent with those predicted theoretically: $\alpha_1\approx 46^{\circ}$ and  $\alpha_2 = \alpha_3\approx 68^{\circ}$. This configuration corresponds to an isosceles triangle with the lower angle placed at the $+1$ defect. However, the defects are not yet in a great circle. In fact, for $h/R= 0.03$, the defects all lie in the upper hemisphere of the shell and are contained in a plane that is perpendicular to the gravitational direction, $\bar{g}$; the angle $\theta_z$ between the normal of this plane, $\bar{N}$, and $\bar{g}$ is zero. As $h/R$ becomes smaller than 0.03, $\theta_z$ increases [see Fig. \ref{fig:experiments1}(e)] and the defects progressively separate from each other while maintaing the values of the surface angles and hence the shape of the isosceles triangle. As this happens, the shell progressively approaches the expected configuration for an infinitely thin nematic shell, where the defects lie on a great circle. Indeed, for $h/R \approx 0.01$, the $+1$ defect is located at the top of the shell, see Fig.\ref{fig:experiments2}(a), while the $+1/2$ defects are in the lower hemisphere of the shell, as indicated with the arrows in Fig.\ref{fig:experiments2}(b) and (c). In this configuration, the distributions for the anbles $\beta_{12}$ and $\beta_{23}$ are both Gaussians chacterized by a mean of $\beta_{12}=118^{\circ}$ and $\beta_{23}=77^{\circ}$ and correspoding widths of $\Delta\beta_{12}=19^{\circ}$ and $\Delta\beta_{23}=15^{\circ}$, as shown in Figs. \ref{fig:experiments2}(d) and (e). These values are slightly lower than the theoretical ones, indicating that the defects have not completely reached yet a great circle, possibly due to a remaining thickness inhomogeneity in the shell. Finally, note that throughout this process the pair of boojums is located in the thinnest hemisphere, whereas in the simulations of ref. \cite{PhysRevE.88.012508} the pair of boojums was found to be located in the thickest hemisphere. These simulations, however, are carried out for shells with a much larger thickness than the shells in this experiment.

\section{Valence transitions}

We will now proceed with an estimate of the energy of the trivalent shell when this escape is taken into account. In doing so, we follow the arguments in ref. \cite{2006PhRvE..74b1711V}. We first consider the energy when three singular lines are spanning the shell at angular distances reported above. We estimate this energy as the product of the two-dimensional result and the thickness, $h$, thus effectively taking $k=Kh$:

\begin{equation}
E'_{Z=3} = \pi K h \left[ \left( 1 + 2 \times \frac{1}{4}\right)\log \left( \frac{R}{a}\right) - 0.54 +\frac{3E_c}{\pi K h}\right]
\end{equation}

A heuristic yet adequate method to include the escape is by replacing the microscopic cut-off by the thickness of the shell, since the singular core is spread out over spatial dimensions of the order of $h$. To account for the pair of boojums an energy $4.2K\pi h$ is added\cite{2002PhRvE..66c0701C,2006PhRvE..74b1711V,C3SM27671F}. We then obtain

\begin{equation}
E_{Z=3} = \pi K h \left[ \log \left( \frac{R}{h}\right) + \frac{1}{2} \log \left( \frac{R}{a}\right) + 3.65 +\frac{2E_c}{\pi K h}\right]
\end{equation} 

By comparing this to the energy of a shell with four disclination lines

\begin{equation}
E_{Z=4} = \pi K h \left[ \log \left( \frac{R}{a}\right) - 0.43 + \frac{4E_c}{\pi K h}\right]
\end{equation}

\noindent we can find the critical value for $h$ above which the trivalent defect configuration is energetically preferable over the tetravalent one:

\begin{equation}
h^*_{34}/R = e^{4.08-2E_c/ \pi K h}\sqrt{\frac{a}{R}}
\end{equation}

Similarly, one can find the critical value for $h$ below which the trivalent defect configuration is energetically preferable over the divalent one by setting $E_{Z=3}$ equal to the approximation of the energy of a shell with two diametrically opposite pairs of surface defects, $E_{Z=2}$. Again, we will first find the energy of a shell with two singular lines

\begin{equation}
E'_{Z=2} = \pi K h \left[ 2\log \left( \frac{R}{a}\right) - 0.69 +\frac{2E_c}{\pi K h}\right]
\end{equation}

\noindent after which we apply the same trick as we used to find $E_{Z=3}$ to obtain

\begin{equation}
E_{Z=2} = \pi K h \left[ 2 \log \left( \frac{R}{h}\right) + 7.69\right]. 
\end{equation}

We find a very similar value

\begin{equation}
h^*_{23}/R = e^{4.04-2E_c/ \pi K h} \sqrt{\frac{a}{R}}
\end{equation}

\begin{figure}[h]
\centering
\includegraphics[width=\columnwidth]{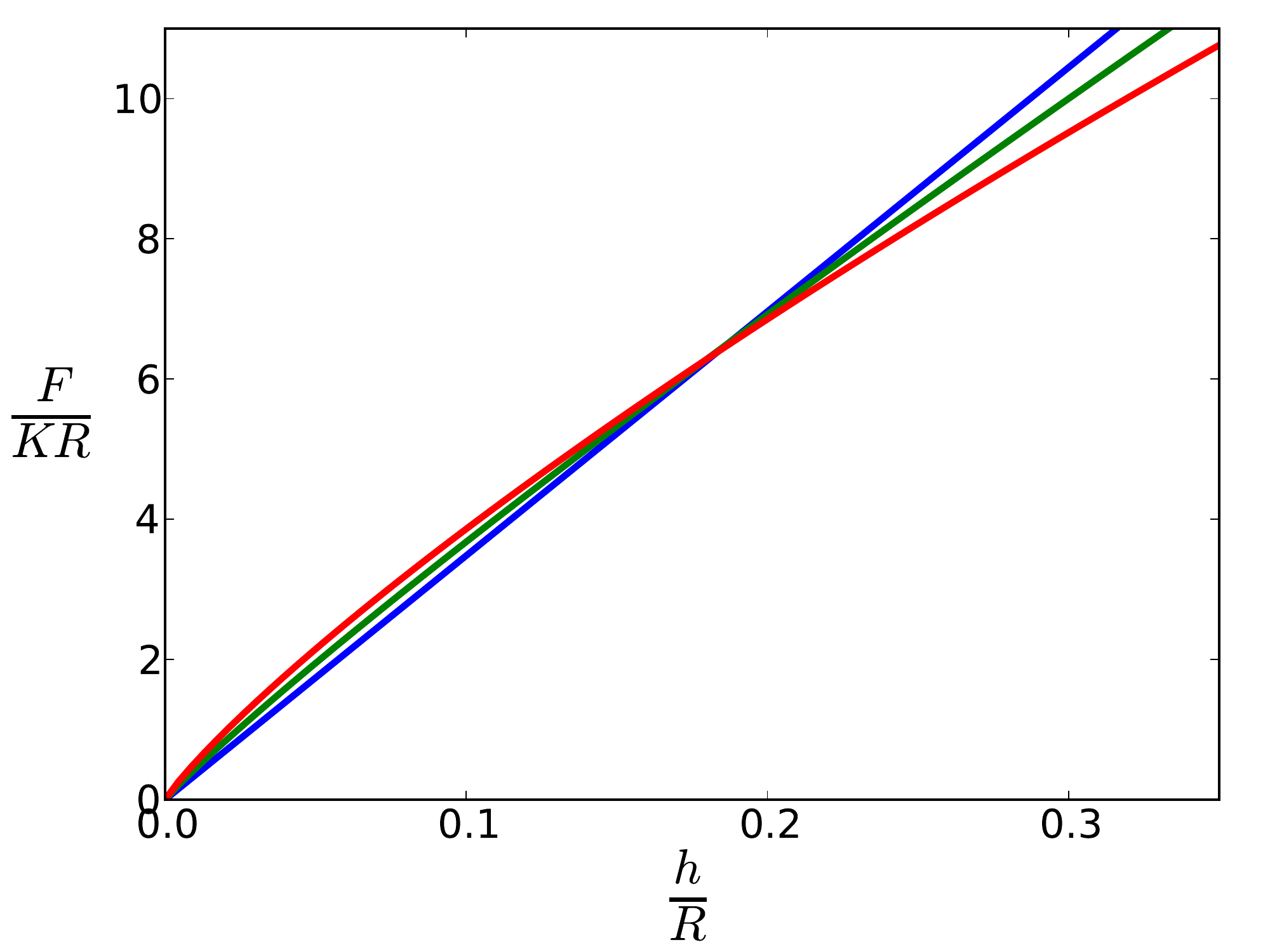} 
\caption{Elastic energy as a function of shell thickness for divalent (red), trivalent (green) and tetravalent (blue) defect configuration for $R/a=10^5$ and $E_c=0$. Either the divalent or tetravalent configuration, but not the trivalent configuration, is lowest in energy.\label{fig:energy_vs_h_all_valence}}
\end{figure}

The energy as a function of thickness is plotted in Fig. \ref{fig:energy_vs_h_all_valence} for all three different valencies. 

 \begin{figure}[h]
\centering
\includegraphics[width=\columnwidth]{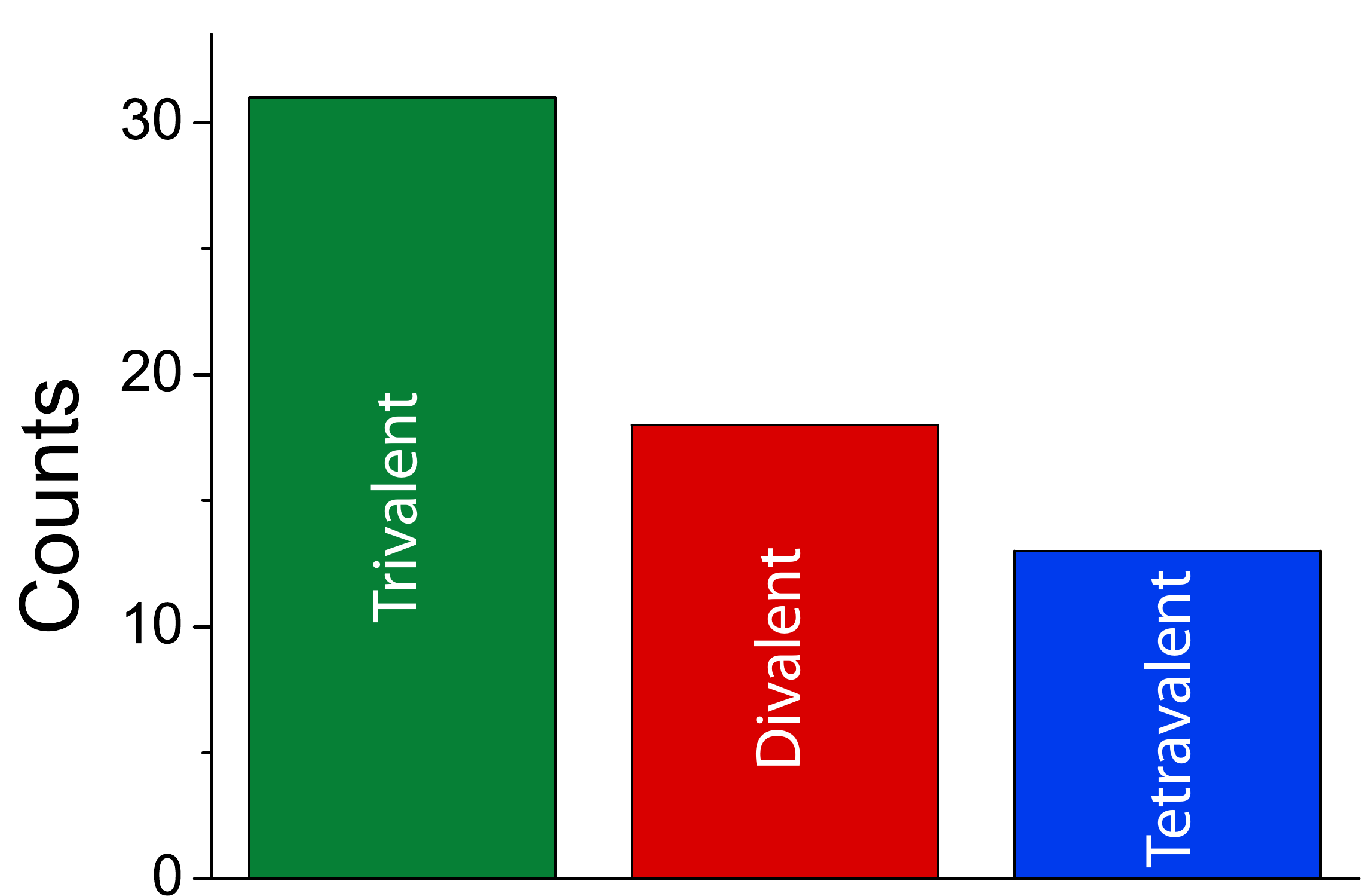} 
\caption{Coexistence of divalent, trivalent and tetravalent configurations in shells with $h/R \simeq 0.1$. The histogram shows the different populations in a sample of 60 shells. \label{fig:experiments3}}
\end{figure}

Since $h^*_{23}<h^*_{34}$ there is no $h$ for which the trivalent shell has lower energy than both the divalent and tetravalent shell. Experimentally, however, we do observe divalent, trivalent and tetravalent shells. Furthermore, they appear with similar likelyhood with a tendency towards formation of the trivalent shell, as shown in the histogram of Fig. \ref{fig:experiments3} for shells with $h/R \simeq 0.1$ made using microfluidics. The presence of all shell types likely result from their similar energies; this makes all three of them accessible while they are being generated. The lack of shell type transitions and hence the observed lifelong stability of a certain shell type irrespective of whether it corresponds or not to the ground state is likely due to energy barrier between shells with different defect number. To ascertain this, we calculate the corresponding energy barriers. For going from the trivalent to the tetravalent shell, we simply have to undo the escape in the third dimension. The associated barrier thus corresponds to the difference between $E'_{Z=3}$ and $E_{Z=3}$:

\begin{equation}
\Delta E_{3\rightarrow4} = \pi K h \left( \log \left( \frac{h}{a} \right) - 4.19 + \frac{E_c}{\pi K h}\right)
\end{equation}

The energy barrier for going from the divalent to the tetravalent shell lies in overcoming the repulsion between the two $+1/2$ defects and can thus be estimated by the difference between $E'_{Z=2}$ and $E'_{Z=3}$:

\begin{equation}
\Delta E_{3\rightarrow2} = \pi K h \left( \frac{1}{2} \log \left( \frac{R}{a} \right) - 0.15 -  \frac{E_c}{\pi K h}\right)
\end{equation}

Since $K\approx 10^{-11} N$, for a thin shell with $h=1 \mu m$, these barriers are four orders of magnitude larger than the thermal energy scale (at room temperature), $k_BT$, where $T$ denotes temperature and $k_B$ is Boltzmann's constant, thus providing stability of any of the shell types after made, explaining why valence transitions are not observed experimentally.

Finally, to understand the tendency towards the trivalent shell, we recall that while the shells are generated, the inner droplet is usually never at the center of the outer liquid crystal droplet. This implies that the shells are heterogeneous in thickness, with a thinner and a thicker part. Since defect nucleation happens at random locations inside the shell and for the thin part $h < h_{23}^*$, where $E_4<E_3<E_2$, we might be providing conditions for formation of two $+1/2$ lines in this region. Similarly, since in the 
thick part of the shell, $h > h_{34}^*$, where $E_2<E_3<E_4$, we might be providing conditions for formation of a pair of $+1$ boojums in this part of the shell. The tendency observed experimentally towards the trivalent shell could then relie on the thickness heterogeneity of the shells as they are made in our microfluidic devices. Once the shells are made, their thickness and thickness inhomogeneity reach their equilibrium configuration and the defects locate at the top, where the elastic free energy of the nematic liquid crystal is minimum.

\section{Bond fidelity}

\begin{figure}[h]
\centering
\includegraphics[width=\columnwidth]{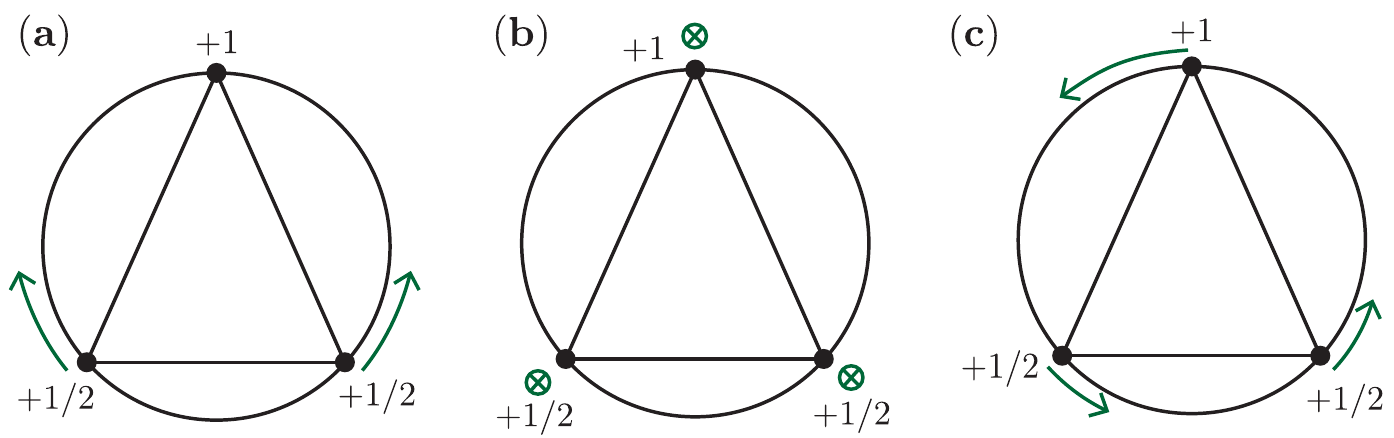} 
\caption{Schematics of the three non-trivial eigenmodes corresponding to (a) $\mathbf{u}_4$, (b) $\mathbf{u}_5$ and (c) $\mathbf{u}_6$. The defects (represented by dots) continue to lie on a  great circle in (a) and (c), but not in (b). The defects continue to form an isosceles triangle in (a) and (b), but not in (c). \label{fig:eigenmodes}}
\end{figure}

In this section, we will consider the fidelity of the three bonds by considering its robustness against thermal fluctuations. We will expand the energy around the equilibrium values for the zenith and azimuthal angles, $\{ \theta^0_i, \phi^0_i \}$. We parametrise the departures from the equilibrium angles with a $2Z$-component vector $\bf{q}$, whose first three components are the deviations along the lines of longitude of the sphere and whose final three components are the deviations along the lines of latitude of the sphere. We thus have 
\begin{align}
q_i = \delta\theta_i, \\
q_{3+i} = \delta\phi_i \sin \theta_i.
\end{align}
Again we employ the law of cosine on the sphere
\begin{align}
\cos \beta_{ij} &= \cos \left( \theta^0_i + q_i \right) \cos \left( \theta^0_j + q_j \right) 
+ \sin \left( \theta^0_i + q_i \right) \sin \left( \theta^0_j + q_j \right) \nonumber \\
&\times \cos \left( \phi^0_i - \phi^0_j + \frac{q_{3+i}}{\sin \theta^0_i} - \frac{q_{3+j}}{\sin \theta^0_j} \right)
\end{align}
to rewrite $F$ in eq. \eqref{eq:free_energy_defect} as an expansion to quadratic order in $q$:
\begin{equation}
F=F\left(\theta^0_i, \phi^0_i\right) + \frac{1}{2}\sum_{ij} M_{ij}  q_i q_j + \mathcal{O}\left(q^4\right). 
\end{equation}
The $6\times6$ matrix $M$ can thus be found by
\begin{equation}
M_{ij} = \left(\frac{\partial^2F}{\partial q_i \partial q_j}\right)_{q_i=q_j=0}
\end{equation}
This calculation is performed without loss of generality upon choosing the ground state defect locations to be on the equator, \textit{i.e.} $\theta^0_i = \pi / 2$. We diagonalise this matrix: 
\begin{equation}
M = U D U^T.
\end{equation}
The matrix $D$ has the following eigenvalues on the diagonal:
\begin{equation}
\{ \lambda_i \} = \frac{\pi k}{20}\{ 0, 0, 0, 15, 17,18\}
\end{equation}
The columns of the matrix $U$ are the corresponding orthonormal eigenvectors, $\{\mathbf{u}_i\}$,  and $U^T$ is the transpose of $U$.
The eigenvectors belonging to the three zero eigenvalues represent rigid body rotations. The other eigenvectors are
\begin{equation}
\mathbf{u}_4 = \begin{pmatrix} 0 \\ 0 \\ 0 \\ 0 \\ -\frac{1}{\sqrt{2}} \\[0.5em] \frac{1}{\sqrt{2}} \end{pmatrix}\!,\
\mathbf{u}_5 = \begin{pmatrix} \frac{4}{\sqrt{34}} \\[0.5em] \frac{3}{\sqrt{34}} \\[0.5em] \frac{3}{\sqrt{34}} \\[0.5em] 0 \\ 0 \\ 0  \end{pmatrix}\!,\
\mathbf{u}_6 = \begin{pmatrix} 0 \\ 0 \\ 0 \\ \sqrt{\frac{2}{3}} \\[0.5em] \frac{1}{\sqrt{6}} \\[0.5em] \frac{1}{\sqrt{6}} \end{pmatrix}\!.
\end{equation}
The fourth and sixth eigenvalues also correspond to deformations that keep the defects located at a great circle. The fourth one corresponds to a displacement of the charge one-half defects such that their distance to the charge one defect grows or shrinks in equal manner and hence preserves the isosceles shape of the triangle (Fig. \ref{fig:eigenmodes}a). 
The sixth eigenvalue corresponds to a mode deformation that does not posses this property, thus breaking the symmetry of reflection of the bisector of the distinct angle (Fig. \ref{fig:eigenmodes}c).  The mode of deformation corresponding to the fifth eigenvalue, however, retains the isosceles shape of the triangle, but shrinks the size of the triangle as the defects do not lie on a great circle anymore (Fig. \ref{fig:eigenmodes}b).
We change the basis from $q_i$ to $w_i$, which is the departure from the trivalent ground state in the $i$-th eigendirection:
\begin{equation}
q_i = U_{ij} w_j.
\end{equation}
This basis transformation yields to quadratic order in $w_i$:
\begin{equation}
F=F\left(\theta^0_i, \phi^0_i\right) + \frac{1}{2} \lambda_4 w_4^2+ \lambda_5 w_5^2+\lambda_6 w_6^2. 
\end{equation}
By equipartition, each term contributes $\frac{1}{2}k_BT$. The eigenvalues corresponding to the trivalent modes of deformation are equal or larger than the the tetravalent ones (which are $\frac{3}{8}\pi k$ and $\frac{3}{4}\pi k$ \cite{2002NanoL...2.1125N, 2006PhRvE..74b1711V}): the trivalent ground state is thus somewhat better protected against thermal fluctuations.

\section{Conclusion}

In a spherical nematic shell of finite thickness a stable defect structure with two $s=1/2$ lines and one pair of boojums is observed experimentally besides the bipolar and regular tetrahedral configuration. The repulsive interdefect interaction pushes the defects to lie on a great circle. The strength of the interaction depends on the charges of the defects. Consequently, the defects are located at the vertices of a isosceles triangle rather than a equilateral triangle, in contrast to the tetravalent ground state in which the defects are equidistant. In the energetically most favourable trivalent configuration, we obtain for the central angles $\beta_{12} = \beta_{13} = 0.73 \pi$, $\beta_{23}=0.54 \pi$ and for the angles in the (flat) isosceles triangle $\alpha_1 = 48^{\circ}$ and $\alpha_2=\alpha_3=66^{\circ}$. These values are in agreement with experimental values. Estimations of the elastic energy show that there is no shell thickness for which the trivalent ground state is lower than both the tetravalent and divalent ground state. However, there are energy barriers to provide stability for the trivalent state once it is created. We note that our calculations do not include thickness heterogeneity. Including this effect is the next step towards our continued understanding of trivalent nematic shells.

\bibliography{liquid_crystals} 

\end{document}